\documentclass[12pt]{iopart}
\usepackage{iopams}  
\usepackage{url}
\usepackage{cite}
\usepackage{lineno}

\begin{document}

\title[]{PSMA PET/CT as a predictive tool for subregional importance estimates in the parotid gland}

\author{\textmd{Caleb Sample}\textsuperscript{\textmd{1,2}}, \textmd{Arman Rahmim}\textsuperscript{\textmd{1,3,4}}, \textmd{Fran\c{c}ois Benard}\textsuperscript{\textmd{3,4,5}}, \textmd{Jonn Wu}\textsuperscript{\textmd{6,7}}, \textmd{Haley Clark}\textsuperscript{\textmd{1,2,7}}}
\vspace{0.5cm}
\address{\textsuperscript{1}Department of Physics and Astronomy, Faculty of Science, University of British Columbia, Vancouver, BC, CA\\
\textsuperscript{2}Department of Medical Physics, BC Cancer, Surrey, BC, CA\\
\textsuperscript{3}Department of Radiology, Faculty of Medicine, University of British Columbia, Vancouver, BC, CA\\
\textsuperscript{4}Department of Integrative Oncology, BC Cancer Research Institute, Vancouver, CA\\
\textsuperscript{5}Department of Molecular Oncology, BC Cancer, Vancouver, BC, CA\\
\textsuperscript{6}Department of Radiation Oncology, BC Cancer, Vancouver, BC, CA\\
\textsuperscript{7}Department of Surgery, Faculty of Medicine, University of British Columbia, Vancouver, BC, CA
}
Corresponding Author and Email: Caleb Sample, csample@phas.ubc.ca

\vspace{0.5cm}

\begin{abstract}

Xerostomia and radiation-induced salivary gland dysfunction remain a common side effect for head-and-neck radiotherapy patients, and attempts have been made to quantify the heterogeneous dose response within parotid glands. Here several models of parotid gland subregional importance are compared with prostate specific membrane antigen (PSMA) positron emission tomography (PET) uptake. PSMA ligands show high concentrations in salivary glands, whose uptake has been previously found to relate to gland functionality. We develop a predictive model for relative importance estimates using PSMA PET and CT radiomic features, and demonstrate a methodology for predicting patient-specific importance deviations from the population. Intra-parotid gland uptake was compared with four regional importance models using 30 [18F]DCFPyL PSMA PET images. A radiomics-based predictive model of population importance was developed using a double cross-validation methodology. Population importance estimates were supplemented using patient-specific radiomic features. Anticorrelative relationships were found to exist between PSMA PET uptake and four independent models of subregional parotid gland importance from the literature. Kernel Ridge Regression with principal component analysis feature selection performed best over test sets (MAE = 0.08), with GLCM features being particularly important. Deblurring PSMA PET images strengthened correlations and improved model performance. This study suggests that regions of relatively low PSMA PET concentration in parotid glands may exhibit relatively high dose-sensitivity. We've demonstrated the utility of PSMA PET radiomic features for predicting relative importance within the parotid glands. PSMA PET appears promising for analyzing salivary gland functionality.

\end{abstract}
\newpage 

\section{Introduction:}
Intensity Modulated Radiotherapy (IMRT) allows for the creation of treatment plans with high dose conformity in cancerous regions while minimizing dose to healthy tissue \cite{powell}. However, high dose levels in cancerous tissue inevitably tail off into healthy tissue, so treatment planners and oncologists must prioritize which healthy regions to spare. Treatment planning in the head-and-neck region is particularly challenging, as there are many organs in close proximity which often abut or overlap with tumour volumes \cite{cheng}. Dose levels in the salivary glands are of particular concern, as xerostomia (subjective sensation of oral dryness) remains a common side effect for head-and-neck cancer patients \cite{ma}. Dose to the largest salivary glands, the parotid glands, is the greatest risk factor for post-treatment xerostomia \cite{chambers}. 

The current standard of care is to minimize the whole-mean dose to parotid glands \cite{deasy}, which were previously considered to have a uniform dose response \cite{eisbruch}. However, there have been numerous attempts in recent years to quantify the relative importance of various parotid gland subregions for predicting post-treatment complications \cite{han,clark_2018, vanluijk_2015, buettner_2012}. 

A potential quantitative imaging method for assessing functionality within salivary glands is prostate-specific membrane antigen (PSMA) positron emission tomography (PET). PSMA PET radioligands target the PSMA, a type 2 integral membrane protein that is expressed in all forms of prostate tissue, including carcinoma \cite{ross2003}. PSMA PET is typically used to stage prostate cancer \cite{Afshar_2015} due to the high relative abundance of PSMA, which increases in proportion to the stage and grade of tumours \cite{silver1997prostate}.
PSMA ligand radiotracers developed for PSMA PET imaging have demonstrated high physiological uptake in the salivary glands, as well as the lacrimal glands, liver, spleen, kidneys, and colon \cite{trover, israeli, wolf,Schwarzenboeck2017}. PSMA PET was furthermore used by Valstar et al. in 2021 \cite{valstar_2021} to discover previously undocumented bilateral salivary glands in the posterior nasopharynx, the ``tubarial glands.'' It is unclear whether uptake in non-prostate tissue is mediated by the expression of PSMA in these tissues, or if PSMA ligand uptake is a result of other physiological mechanisms \cite{Schwarzenboeck2017}.

PSMA ligand uptake in salivary glands has been suggested \cite{klein} to correlate with functional capacity, and mounting evidence supports this relationship \cite{klein, zhao, mohan, hotta}. It was also recently shown that PSMA PET in the parotid and submandibular glands decreased exponentially with radiotherapeutic dose received in the glands during radiotherapy \cite{Mohan2022}. The same study showed that decreases in PSMA PET uptake following treatment were correlated with post-treatment xerostomia. Furthermore, uptake within parotid glands has been found to be non-uniform, with high uptake regions tending towards lateral, posterior, and superior regions \cite{sample_2023_hetero}. We hypothesize that intra-parotid gland uptake variability of PSMA PET is predictive of functional importance. 

The purpose of this study is two-fold. First, we use a data set of 30 PSMA PET images to compare intra-parotid PSMA PET uptake trends with several regional importance estimates from the literature \cite{clark_2018, han, vanluijk_2015, buettner_2012}. Second, we develop a population-level model of Clark et al.'s \cite{clark_2018} regional importance using radiomic features from  PSMA PET and Computed Tomography (CT). We also demonstrate how such a model can be used to predict a patient's deviation from population-derived importance estimates, creating a single metric with population-derived, and patient-specific components.

\section{Methods:}

    \subsection{2.1 Dataset}
        This study was approved by an institutional review board. The data set included de-identified [18F]DCFPyL PSMA PET/CT images for 30 previous prostate cancer patients (Mean Age 68, Age Range 45-81; mean weight: 90 kg, weight range 52 - 128 kg). Scans were acquired two hours following intravenous injection, from the thighs to the top of the skull on a GE Discovery MI (DMI) scanner.The mean and standard deviation of the injected dose was $310 \pm 66$ MBq (minimum: 182 MBq, maximum: 442 MBq). PET images were reconstructed using VPFXS (OSEM with time-of-flight and point spread function corrections) (pixel spacing: 2.73 - 3.16 mm, slice thickness: 2.8 - 3.02 mm). The scan duration was 180 s per bed position. 
        
        Helical CT scans were acquired on the same scanner (kVP: 120, pixel spacing: 0.98 mm, slice thickness: 3.75 mm). Images were scaled to standard uptake values normalized by lean body mass ($SUV_{lbm}$). Registered CT images were used for delineating parotid and submandibular glands. Limbus AI \cite{limbus} was used for preliminary auto-segmentation of the glands, which were then manually refined by a single senior radiation oncologist, Jonn Wu.
        
    \subsection{2.2 Correction of Partial Volume Effects}
        One weakness of PET as an imaging modality is its intrinsically low spatial resolution \cite{marquis_2023}. The burden of partial volume effects is less pronounced when analyzing large geometric regions with homogeneous uptake, but cannot be ignored when attempting to compare heterogeneous uptake in small regions-of-interest (ROIs), such as subregions of the parotid glands. Recently, a method has been developed for simultaneous deblurring and supersampling of PSMA PET images using neural blind deconvolution \cite{sample_2023_blind_deconv}, which we employ in this work for preprocessing PSMA PET images. This model has been shown to illuminate fine uptake trends within small regions of parotid glands. We performed all calculations with both "enhanced" images, and unmodified original images. 

    \subsection{2.3 Comparison of PSMA PET with Parotid Gland Importance Models}
    PSMA PET uptake trends were compared with four models of intra-parotid gland importance found in the literature, detailed in the following sub-sections. Uptake metrics included the mean, median, and maximum, calculated in ROIs defined according to each specific model. As importance models were all population-level estimates, uptake metrics were averaged over the 60 parotid glands from the 30 patients.

        \subsubsection{Clark et al.'s Model}
\begin{figure}[h]
              \centering
              \includegraphics[width=1\textwidth]{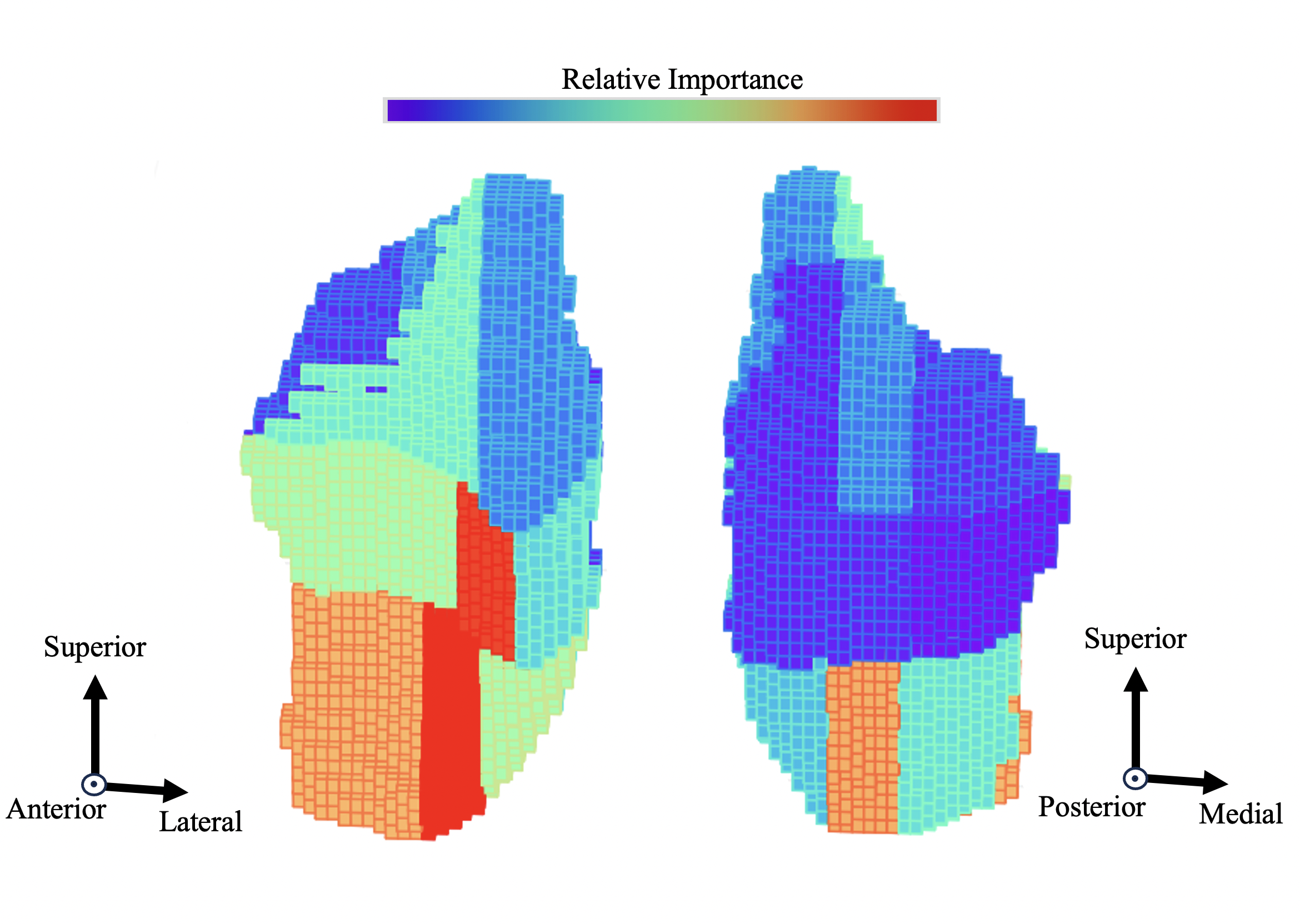}
              \caption{3D rendering of voxels within a parotid gland corresponding to Clark et al's relative importance subregions are shown from two angles.}
              \label{fig:clark_imp}
            \end{figure}

            Clark et al.'s model \cite{clark_2018} estimates the relative importance of 18 equal-volume subregions of the contralateral parotid gland for predicting salivary dysfunction following radiotherapy. Stimulated saliva measurements were collected for 332 patients before and at one year after radiotherapy. The relative differences were predicted with conditional inference trees using radiotherapeutic dose levels in parotid gland subregions. Parotid glands were sub-segmented using nested planar segmentation (planes: 2 axial, 1 coronal, 2 sagittal). Regions of high relative importance tended towards caudal-anterior regions, as shown in Fig~\ref{fig:clark_imp}.
            We sub-segmented parotid glands according to the same regimen, and voxels within each of the 18 subregions were used to calculate uptake statistics. To test whether intra-parotid PSMA PET uptake is related to Clark et al.'s importance estimates, Spearman's rank correlation coefficient, $r_s$, was computed between uptake in the 18 subregions along with their corresponding importance estimates.

        \subsubsection{Han et al.'s model} 
\begin{figure}[h]
              \centering
              \includegraphics[width=1\textwidth]{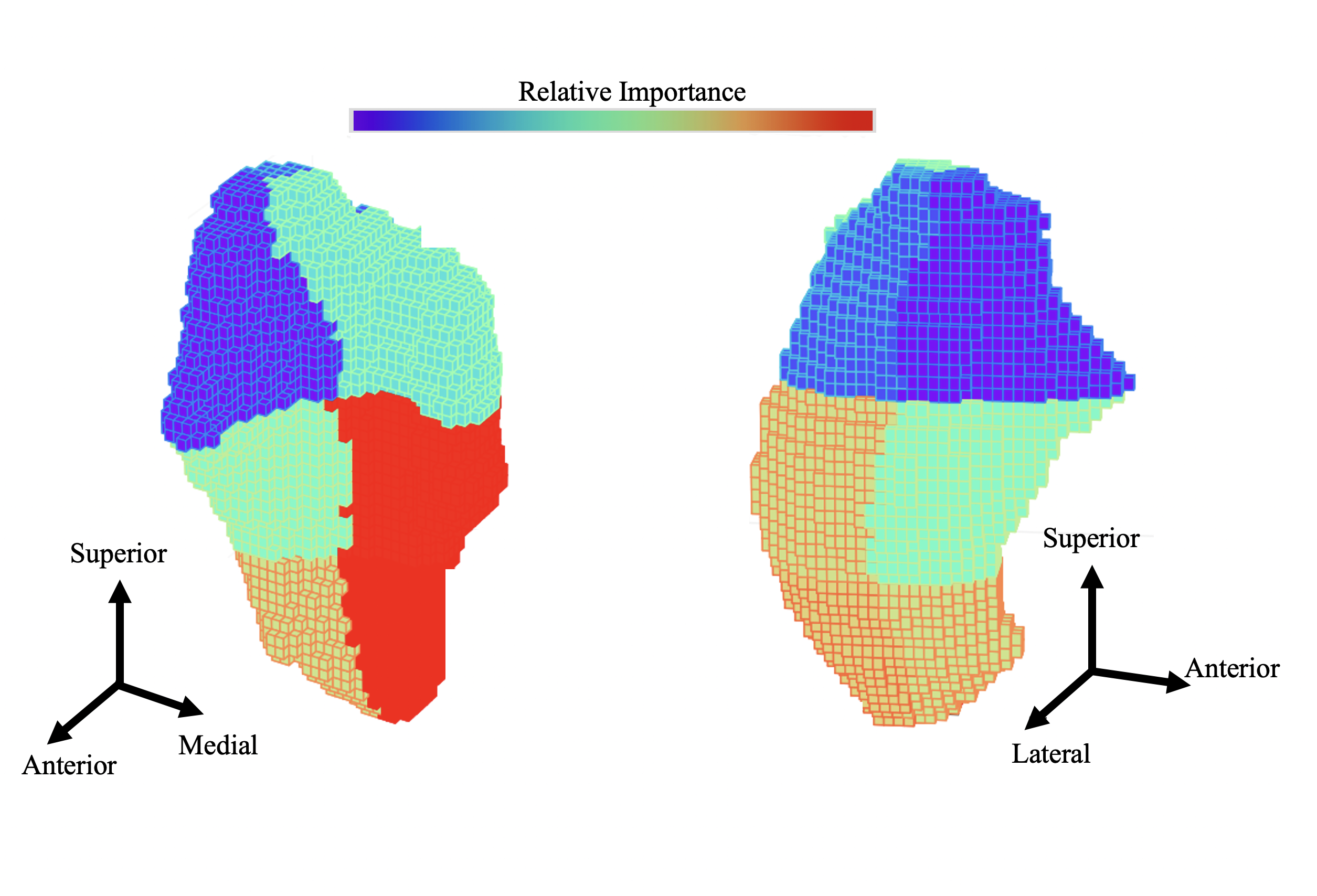}
              \caption{3D rendering of voxels within a parotid gland corresponding to Han et al.'s relative importance subregions are shown from two angles.}
              \label{fig:han_imp}
            \end{figure}

            Han et al. \cite{han} assess the relative importance of 9 parotid gland subregions for predicting injury ($\ge$ grade 2 xerostomia at 6 months post-radiotherapy) and recovery ($\ge$ grade 2 xerostomia at 6 months post-radiotherapy, followed by $<$ grade 2 xerostomia at 18 months post-radiotherapy). Subregions were defined by first applying a 3 mm margin to whole parotid glands, then dividing glands into three radial sectors (anterior, medial, posterior), then further dividing these sectors along the inferior-superior axis into 3 equal-length regions. Voxels within a parotid gland corresponding to these importance regions are shown in Fig~\ref{fig:han_imp}.
            Han et al. \cite{han} determined the relative importance of 9 dose-volume statistics in 10\% volume increments from D10 (Minimum dose to 10\% volume) to D90 (Minimum dose to 90\% volume) in each subregion. For our purposes, the mean importance computed over all dose statistics was used as a single relative importance estimate for each subregion. Spearman's rank correlation coefficient, $r_s$, was calculated between uptake and relative importance for predicting injury, and recovery.

        \subsubsection{Van Luijk et al.'s Model}
\begin{figure}[h]
              \centering
              \includegraphics[width=0.9\textwidth]{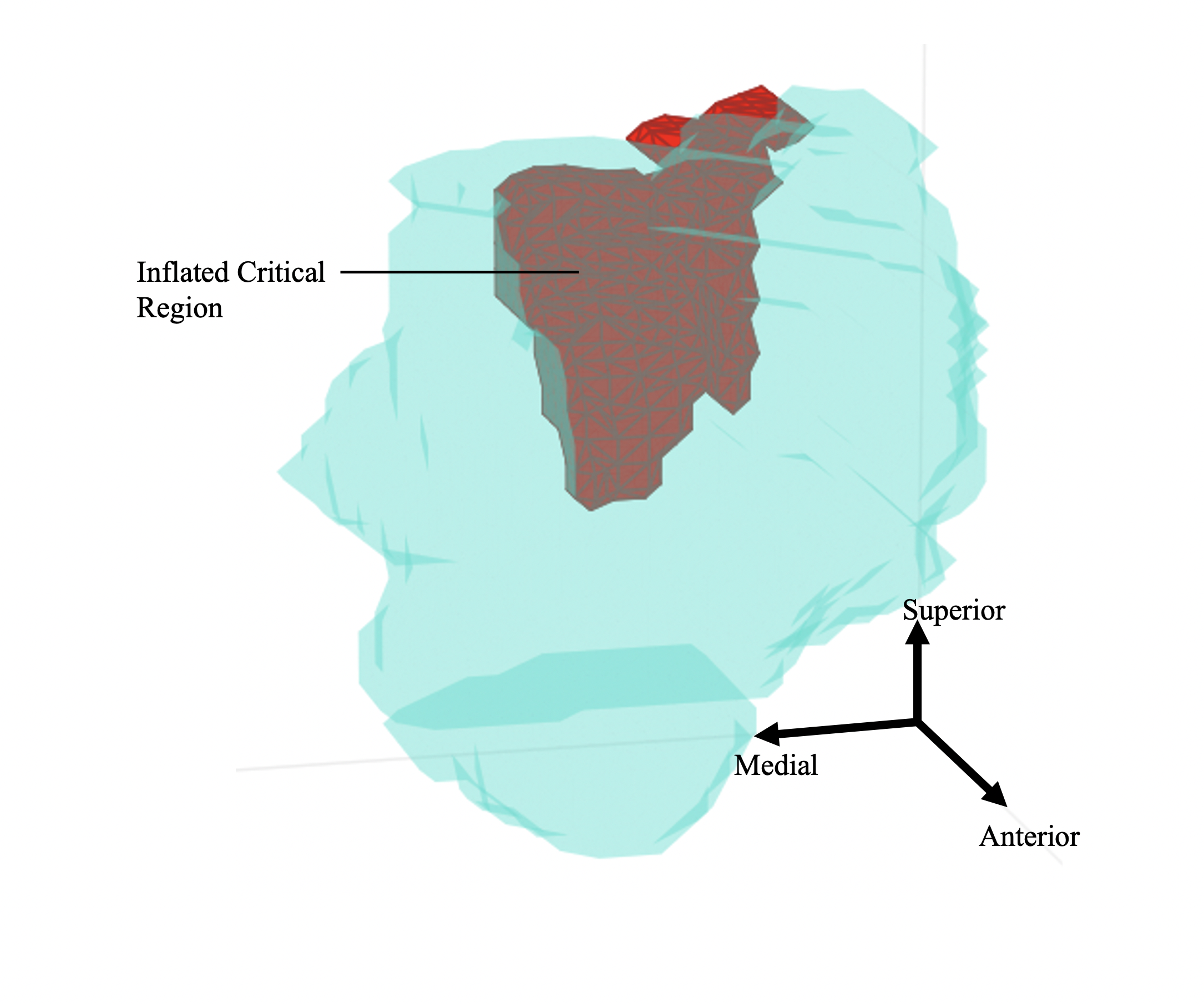}
              \caption{The approximate location of Van Luijk et al.'s critical region of the parotid gland used for computing uptake statistics is shown}
              \label{fig:vanluijk}
        \end{figure}

        Van Luijk et al. \cite{vanluijk_2015} used stimulated saliva measurements and radiotherapeutic dose levels to locate "critical" regions within parotid glands which are most predictive of salivary outcome at one year post radiotherapy. The study did not specify a well-defined location of this critical region over the population, however, it is stated to be in close proximity of the Stensen's duct, adjacent to the dorsal side of the mandible. For our purposes, we approximated the critical region by applying a 9 mm margin to the mandible, which was intersected with the top half of the parotid gland (Fig~\ref{fig:vanluijk}). The 9 mm margin was found to consistently intersect a region of the parotid gland  approximately corresponding Van Luiijk et al.'s \cite{vanluijk_2015} critical regions.
        Uptake statistics were compared within expanded critical and non-critical regions using a paired t-test.
        
        \subsubsection{Buettner et al.'s Model}
        
        Buettner et al. \cite{buettner_2012} evaluated the predictive ability of various dose "moments" in a regression model for post-treatment xerostomia in 63 head-and-neck cancer patients, treated with either IMRT or conventional radiotherapy. Important variables included mean dose to the superficial lobe, skewness of dose in the cranial-caudal direction within the deep and superficial lobe, and relative concentration of dose in the caudal-medial region of the deep lobe. While the parotid glands were only segmented into superficial and deep lobes, dose moments calculated within these regions evaluated the spatial variance of the dose response.
            
        \subsection{2.4 Development of a predictive Model for parotid gland relative importance using PSMA PET and CT}
        To demonstrate the predictive ability of PSMA PET radiomic features for parotid gland regional functionality, we develop a model for predicting Clark et al.'s \cite{clark_2018} relative importance estimates using radiomic features extracted from both PSMA PET and CT images. We then demonstrate how such a model can be used for predicting patient-specific perturbations away from population-level importance estimates.

        To explicitly demonstrate the benefit of model-building using radiomic features versus using only standard uptake statistics (mean, median, maximum), we also develop another model using the same methods to be described but using only the mean, median, and maximum uptake as input features.
        
        \subsubsection{Feature Extraction}
        The standardized pyradiomics library \cite{pyradiomics} was used for computing radiomic features of PSMA PET and CT within all 18 of Clark's subregions. The full set of Gray Level Co-occurrence Matrix (GLCM), Gray Level Run Length Matrix (GLRLM), Gray Level Size Zone Matrix (GLSZM), Gray Level Dependence Matrix (GLDM), and first order features, computed with original squared, square root, and wavelet image types, were computed for a total of 1060 features. For gray level discretization, a fixed bin width was chosen over a fixed bin count, as a fixed bin width has been shown to have better reproducibility \cite{binning}, especially when chosen to yield a bin count between 16-128 \cite{binning_2}. We therefore set the bin count to 0.2 for original, 0.1 for square root, and 1 for square. For the wavelet features, a fixed bin count of 100 was used, due to uncertainty in the expected range. 
    
        Radiomic features were calculated for individual patients, and averaged over each parotid gland for all 18 patients. This yielded a population-level design matrix of shape (36, 1060) prior to feature selection. Features were calculated for both enhanced and original PSMA PET images.
    
        \subsubsection{Double Cross Validation}
\begin{figure}[h]
                  \centering
                  \includegraphics[height=0.9\textheight]{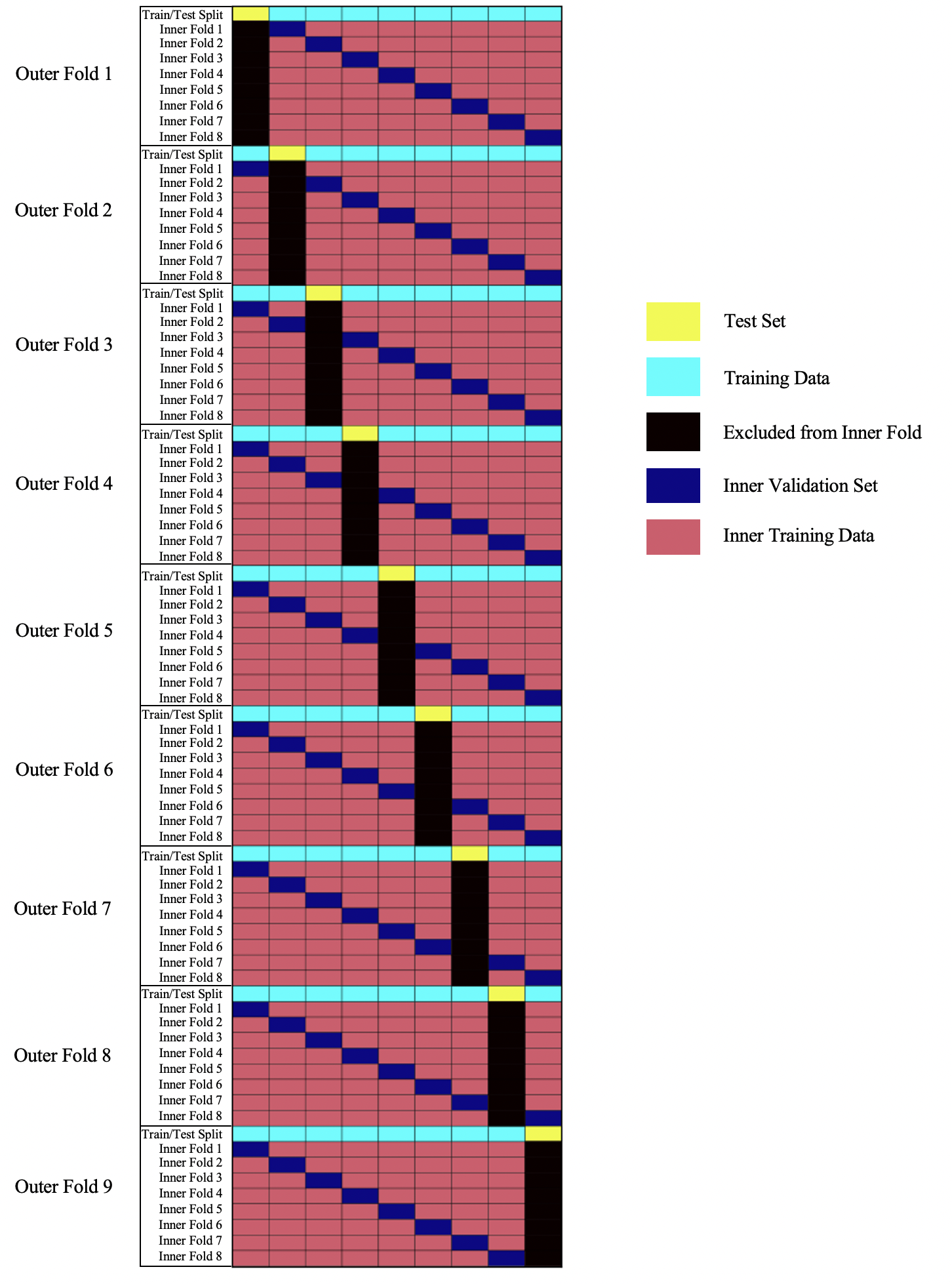}
                  \caption{For model testing and validation, a double cross validation scheme was employed, where the outside test set is rotated through a 9-fold cross validation loop, each including its own 8-fold inner cross validation loop for parameter tuning.}
                  \label{fig:cv}
            \end{figure}

            The small size of our data set made it inappropriate to define a single test set for final performance evaluation, and we therefore used double cross validation, or sometimes called nested cross validation, where our test set was rotated through 9 outside folds, each having its own inner cross validation loop for tuning the feature selection algorithm, model, and hyper-parameters (Fig~\ref{fig:cv}). 
    
        \subsubsection{Feature Selection and Models}
            To avoid overfitting, the large number of features extracted must be pruned using feature selection methods prior to model training. For this purpose, we include 3 feature selection algorithms within the cross validation loops, including a linear combination filter, a pairwise correlation filter, and principle component analysis as used by Delzell et al. \cite{delzell} to predict lung cancer using radiomic features.  The linear combination (lincom) filter uses QR decomposition to iteratively remove features which are linear combinations of others. The pairwise correlation filter tests the correlation between features and removes those who are correlated above a specified cutoff. Principal component analysis changes the basis of the feature space to capture a large portion of the variance using a smaller number of feature vectors. For more information, refer to the work by Delzell et al. \cite{delzell}
       
            Five different regression model types were included within cross validation. This included a linear regression, support vector machine, random forest, conditional inference tree, and kernel ridge model. Model performance is highly dependent on a variety of hyper-parameters which can be tuned for the different models and feature selection algorithms. A set of different hyper-parameters for each model were iterated over within cross validation. All models, feature selection algorithms, and their corresponding hyper-parameters tested are listed in Table~\ref{tab:models_and_fs_tried} Models were scored according to the mean absolute error (MAE).

            %List of tried models / features table
            \definecolor{Dark_Gray}{rgb}{0.5,0.5,0.5}
            \definecolor{Gray}{rgb}{0.7,0.7,0.7}
            \definecolor{whiter}{rgb}{0.8,0.8,0.8}
            \definecolor{white}{rgb}{1,1,1}
            \newcolumntype{a}{>{\columncolor{Gray}}c}
            \newcolumntype{b}{>{\columncolor{whiter}}c}
            \newcolumntype{d}{>{\columncolor{white}}c}
            \begin{table}[h]
                    \centering
                    \captionsetup{justification=raggedright}
                    \caption{Models and feature selection (F.S) algorithms, along with their corresponding hyper-parameters tested in cross-validation, are shown.}
                    \begin{tabular}{@{}cc}
                    \br
                    \rowcolor{white}Model&Hyper-parameters\\
                    \mr
                    \rowcolor{white}Support Vector Machine&$\epsilon = [0.01,0.05,0.1]$\\
                    \rowcolor{white}&Kernel = [Linear, radial basis function, sigmoid, poly]\\
                    \rowcolor{white}&Degree = [2,3]\\
                    \rowcolor{white}&$\gamma = $ [scale, auto]\\
                    \rowcolor{white}&coef0 = $[-0.3, -0.2, -0.1, 0]$\\
                    Random Forest&number estimators $= [3,5,7,10]$\\
                    &max depth $= [3,5,8, None]$\\
                    &Criterion = [Absolute Error, Squared Error]\\
                    \rowcolor{white}Conditional Inference Trees&Max Depth $= [5,10,15,20,25,None]$\\
                    \rowcolor{white}&Criterion = [Absolute Error, Squared Error]\\
                    Kernel Ridge&$\alpha = [0.1, 0.5, 1, 5, 10]$\\
                    &Kernel = [Linear, Radial Basis Function, Sigmoid]\\
                    \rowcolor{white}Linear Regression&N/A\\
                    \br

                    \rowcolor{white}F.S. Algorithm&Hyper-parameters\\
                    \mr
                    Pairwise Correlation Filter&feature count = [1,2,3,4,5,6], cutoff = [0.85, 0.88, 0.9, 0.92]\\
                    \rowcolor{white}PCA&feature count = [1,2,3,4,5,6,8,10,15,20,30]\\
                    Linear Combination Filter&correlation cutoff = [0.05, 0.1, 0.2, 0.3]\\
                    \br
                \end{tabular}\\
                \label{tab:models_and_fs_tried}
            
            \end{table}
        \subsubsection{Error Analysis}
            For estimating the uncertainty of model predictions, we employ the methodology described by Cawley et al. \cite{error_analysis} for kernel ridge models. This involves computing the leave-one-out absolute error for each sample in the data set, and then training a second kernel ridge model for predicting the absolute error of predictions based on the same input features and using this as an estimate of prediction variance.

        \subsubsection{Demonstrating a method for predicting patient-specific importance perturbations}
            Finally, we demonstrate how patient-specific deviations from population-level importance estimates can be obtained, to create a single, combined importance estimate including both population-level and patient-specific components. This is obtained by first computing and processing a patient's radiomic features according to the feature selection algorithm employed by the final model. As the model has been trained to understand the relationship between specific features and relative importance estimates, inputting patient specific features into the model for all 18 parotid gland subregions provides an estimate of relative importance for said patient. A single combined importance estimate is created by taking the population level subregion importance estimates, $I^P_j$, $j \in \mathbb{Z}, 1 \leq j \leq 18$ and the difference in patient specific and population estimates, $\Delta_j$, $j \in \mathbb{Z}, 1 \leq j \leq 18$ and computing
            \begin{equation}
                I = \begin{cases}
                I^P_j & \Delta_j < 0 \\
                \frac{2I^P_j}{1+e^{-2\Delta_j}} &\Delta_j > 0 \\
                \end{cases}
            \end{equation}
            This defines a minimum importance estimate using the population-level estimates, and increases estimates in regions of high patient-specific importance, levelling off as the patient-specific estimate approaches about 3x the maximum population-level importance estimate. Using this approach, patient-specific predictions can be used to supplement or perturb population-level estimates in regions predicted to be of high radiotherapeutic importance. Combined importance estimates for subregions are never lower than population-level estimates, to avoid potential negative impacts associated with under-estimating importance.

\section{Results:} 
    \subsection{3.1 Comparison of PSMA PET with Importance Models}
        Overall, uptake of PSMA PET was found to be inversely proportional with subregion importance estimates from the literature. These trends appeared stronger when enhanced images were used for uptake calculations.
    
\begin{figure}[H]
          \centering
          \includegraphics[width=0.92\textwidth]{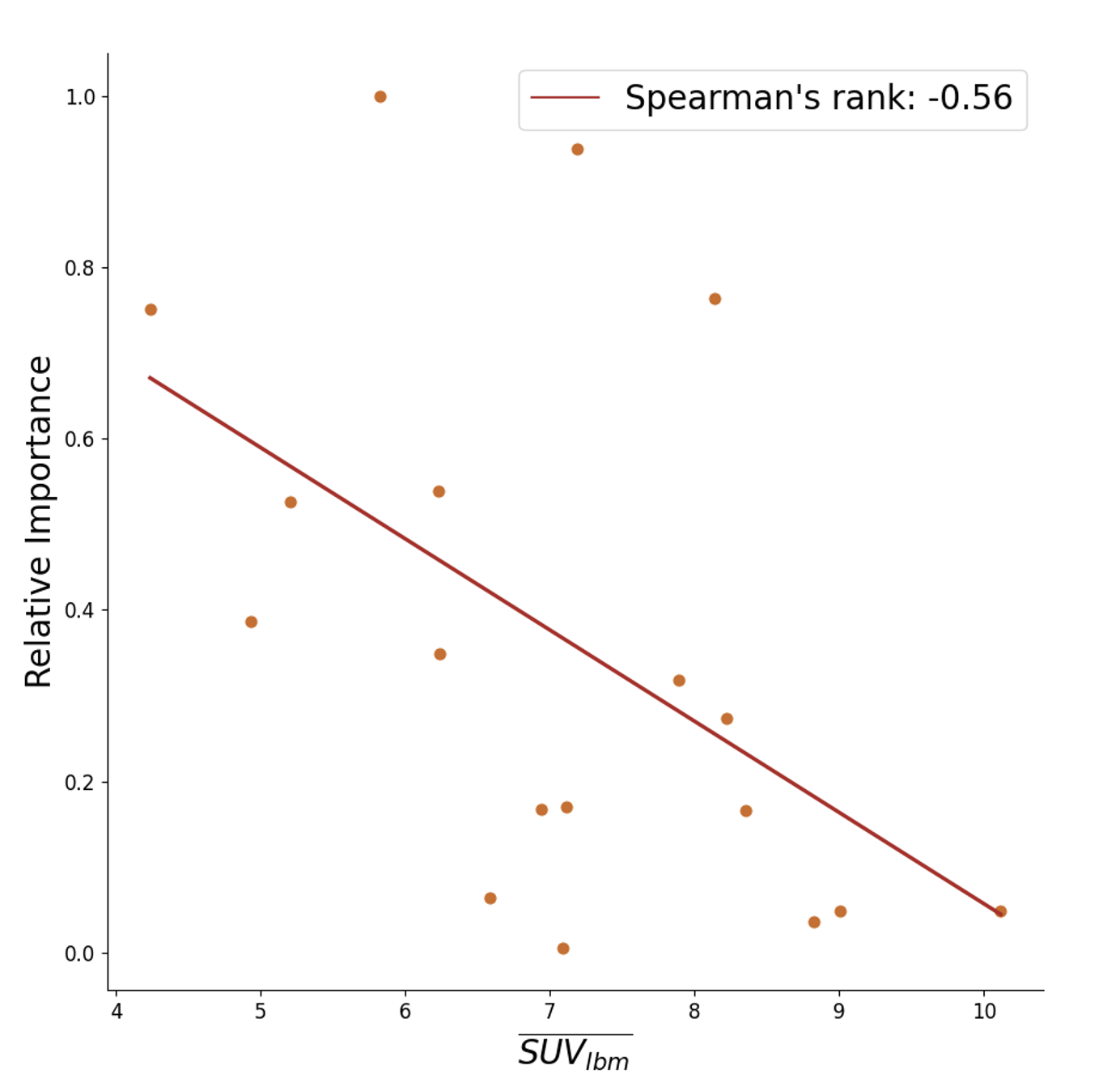}
          \caption{Clark's relative importance vs mean PSMA PET uptake in 18 equal-volume parotid gland subregions, averaged over 30 patients. Relative importance was found to have a significant ($p=0.015$) anti-correlation with regional PSMA PET uptake. Calculations were performed with deblurred PSMA PET images. A best fit line is shown in red. }
          \label{fig:Clark_Scatter}
        \end{figure}

        %Clark
        Clark et al.'s \cite{clark_2018} importance predictions in the 18 equal volume regions were significantly anti-correlated with mean and median uptake (Table~\ref{tab:clark}). A scatter plot of importance vs mean uptake in Clark et al.'s subregions is shown in Fig~\ref{fig:Clark_Scatter}.

    %clark scatter plot
\begin{figure}[H]
          \centering
          \includegraphics[width=0.77\textwidth]{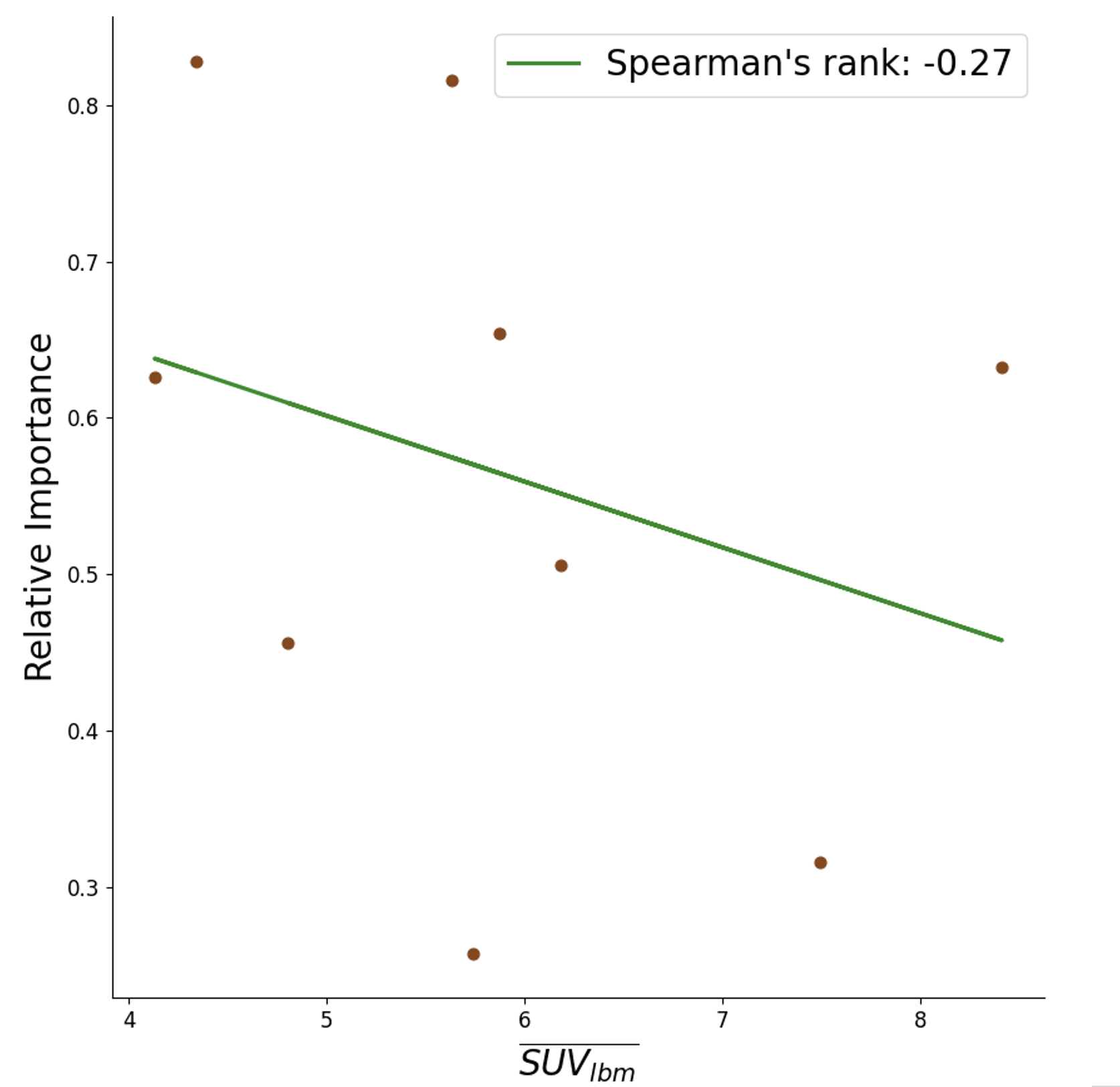}
          \caption{Han's relative importance vs mean PSMA PET uptake in nine parotid gland subregions, averaged over 30 patients. Relative importance was found to have a statistically insignificant ($p=0.015$) anti-correlation with regional PSMA PET uptake. Calculations were performed with deblurred PSMA PET images. A best fit line is shown in green. }
          \label{fig:han_scatter}
        \end{figure}

\begin{figure}[H]
          \centering
          \includegraphics[width=0.8\textwidth]{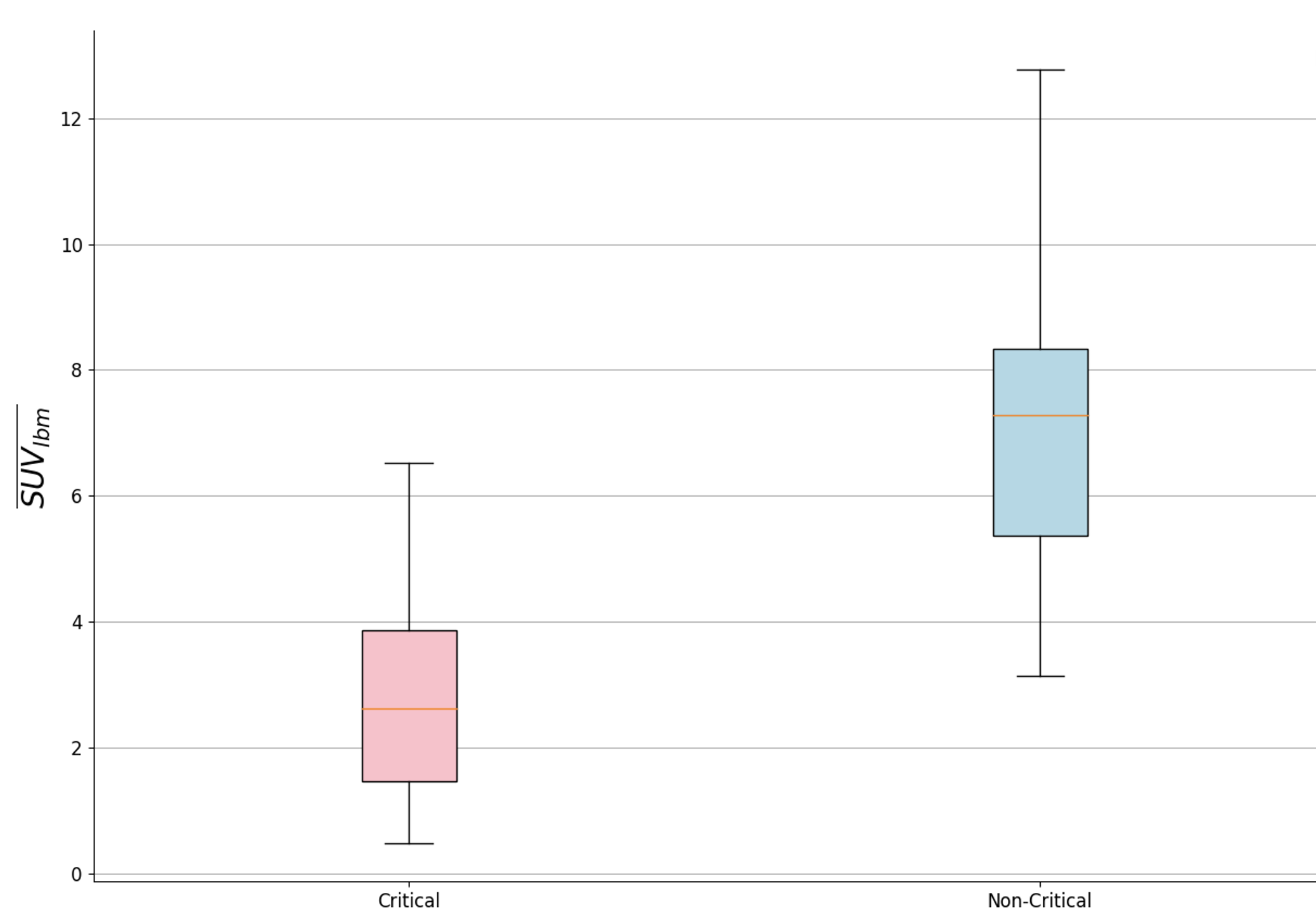}
          \caption{Box plot of the mean PSMA PET uptake in critical and non-critical regions of parotid glands, as defined by Van Luijk et al \cite{vanluijk_2015}. Results are calculated using a dataset of 30 patients. Mean uptake in critical regions was found to be significantly (p $<$ 0.01) lower than in non-critical regions. Calculations were performed with deblurred PSMA PET images.}
          \label{fig:Van_Luijk_Box}
        \end{figure}

            %Clark Rel importance table
        \begin{table}[h]
            \centering
            \captionsetup{justification=raggedright}
            \caption{Spearman's rank correlation coefficients for PSMA PET uptake with Clark's relative importance estimates. Correlations are calculated for mean, median, and maximum uptake, normalized by lean body mass. Results are calculated using enhanced (deblurred and supersampled) and original PSMA PET images.}
            \footnotesize
            \begin{tabular}{@{}cccccc}
            \br
            &Mean&Median&Maximum\\
            \mr
            Enhanced&$r_s = -0.56, p = 0.015$&$r_s = -0.55, p = 0.016$&$r_s = -0.25, p = 0.31$\\
            Original&$r_s = -0.50, p = 0.03$&$r_s = -0.51, p = 0.03$&$r_s = -0.30, p = 0.22$\\    
            
            \br
            \end{tabular}\\
            \label{tab:clark}
    
        \end{table}
        %Han
        Han et al's \cite{han} model predictions for relative importance in 9 regions of unequal volume were not significantly correlated with uptake metrics. There did however exist weak trends of anti-correlation of uptake with importance for predicting injury, and direct correlation for predicting recovery (Table~\ref{tab:han}). A scatter plot of importance vs mean uptake in Han et al.'s subregions is shown in Fig~\ref{fig:han_scatter}.
        Uptake levels were found to be approximately two times higher in Van-Luijk et al.'s \cite{vanluijk_2015} non-critical region than in the critical region ($p < 0.01$) (Table~\ref{tab:vanluijk}). A boxplot of mean uptake in critical and non-critical regions is shown in Figure~\ref{fig:Van_Luijk_Box}

        %Han Table
        \begin{table}[H]
            \centering
            \captionsetup{justification=raggedright}
            \caption{Spearman's rank correlation coefficients for PSMA PET uptake with Han's relative importance estimates for predicting both injury, and recovery. Correlations are calculated for mean, median, and maximum uptake, normalized by lean body mass, using enhanced (deblurred and supersampled) and original PSMA PET images.}
            \footnotesize
            \begin{tabular}{@{}ccccccc}
            \br
            &\multicolumn{2}{c}{$r_s, p$ (Mean \& Imp)}&\multicolumn{2}{c}{$r_s, p$ (Median \& Imp)}&\multicolumn{2}{c}{$r_s, p$ (Maximum \& Imp)}\\
            &Injury&Recovery&Injury&Recovery&Injury&Recovery\\
            \mr
            Enhanced&-0.27, 0.49&0.39, 0.29&-0.07, 0.86&0.12, 0.76&-0.56, 0.11&0.49, 0.18\\
            Original&-0.41, 0.26&0.42, 0.26&-0.19, 0.61&0.18, 0.64&-0.56, 0.11&0.49, 0.18\\
            
            \br
            \end{tabular}\\
            \label{tab:han}
    
        \end{table}

        %Van Luijk Table
        \begin{table}[H]
            \centering
            \captionsetup{justification=raggedright}
            \caption{Comparisons for mean, median, and maximum uptake, normalized by lean body mass, in Van Luijk et al.'s \cite{vanluijk_2015} critical and non-critical parotid gland subregions. Calculations are shown using enhanced (deblurred and supersampled) and original PSMA PET images.}
            \footnotesize
            \begin{tabular}{@{}ccccccc}
            \br
            &\multicolumn{2}{c}{Mean}&\multicolumn{2}{c}{Median}&\multicolumn{2}{c}{Maximum}\\
            &Critical&Non-Critical&Critical&Non-Critical&Critical&Non-Critical\\
            \mr
            Enhanced&$3.1 \pm 2.1$&$7.2 \pm 2.1$&$2.7 \pm 2.1$&$7.2 \pm 2.2$&$9.4 \pm 5.1$&$20.0 \pm 6.0$\\
            Original&$3.2 \pm 2.0$&$6.3 \pm 1.7$&$2.9 \pm 2.1$&$6.2 \pm 1.8$&$7.6 \pm 4.0$&$15.6 \pm 4.5$\\
            
            \br
        \end{tabular}\\
        \label{tab:vanluijk}

    \end{table}
        Uptake statistics in parotid gland subregions corresponding to Buettner et al.'s important regions \cite{buettner_2012} are shown in Table~\ref{tab:buettner}. As shown previously \cite{sample_2023_hetero}, uptake in parotid glands appears skewed towards lateral, posterior, and superior regions. Buettner et al. found the superficial and the deep lobe to be important, with importance concentrating slightly towards caudal regions. The caudal-medial subregion of the deep lobe, which was predicted to be of high importance, was found to have significantly lower (p $<$ 0.01) uptake than in the caudal-lateral subregion. Caudal regions, in general, tend to have lower uptake than superior regions of the glands.

%Buettner table
\definecolor{Gray}{rgb}{0.5,0.5,0.5}
\definecolor{whiter}{rgb}{0.8,0.8,0.8}
\definecolor{white}{rgb}{1,1,1}
\newcolumntype{a}{>{\columncolor{Gray}}c}
\newcolumntype{b}{>{\columncolor{whiter}}c}
\newcolumntype{d}{>{\columncolor{white}}c}
\begin{table}[H]
        \centering
        \captionsetup{justification=raggedright}
         \caption{Buettner et al. \cite{buettner_2012} found that dose to the superficial lobe, relative concentration of dose in the caudal/cranial region of both superficial and deep lobes, and relative concentration in the caudal-medial region of the deep lobe, were predictive of post-treatment xerostomia for head-and-neck radiotherapy patients. It is unclear whether xerostomia is directly or inversely proportional to these metrics, so we simply report differences in PSMA uptake within corresponding regions. Correlations are calculated for mean, median, and maximum uptake, normalized by lean body mass, using enhanced (deblurred and supersampled) and original PSMA PET images. Corresponding subregion pairs are shown side-by-side using the same shading.}
        \tiny
        \begin{tabular}{@{}ddddbbddbb}
        \br
        \rowcolor{white}&&Sup&Deep&Sup Cranial&Sup Caudal&Deep Cranial&Deep Caudal&Deep Caudal-Medial&Deep Caudal-Lateral\\
        \mr
        Enhanced&Mean&$7.4 \pm 2.2$&$5.7 \pm 2.0$&$8.4 \pm 2.7$&$7.2 \pm 2.1$&$6.2 \pm 2.0$&$5.1 \pm 2.2$&$4.2 \pm 2.3$&$5.9 \pm 2.6$\\
        &Medial&$7.6 \pm 2.4$&$5.3 \pm 2.3$&$8.8 \pm 2.8$&$7.0 \pm 2.2$&$5.8 \pm 2.3$&$4.6 \pm 2.6$&$5.7 \pm 2.6$&$5.7 \pm 3.0$\\
        &Maximum&$19.8 \pm 6.0$&$16.3 \pm 4.7$&$18.8 \pm 5.6$&$18.3 \pm 5.8$&$15.7 \pm 4.6$&$13.8 \pm 4.6$&$14.8 \pm 4.7$&$15.2 \pm 4.2$\\
        Original&Mean&$6.5 \pm 1.9$ & $5.2 \pm 1.7$ & $8.0 \pm 2.5$ & $6.3 \pm 1.8$ & $5.6 \pm 1.9$ & $4.5 \pm 1.9$ & $3.5 \pm 1.9$&$5.4 \pm 2.3$\\
        &Median&$6.5 \pm 1.9$ & $5.1 \pm 1.9$ & $8.4 \pm 2.7$ & $6.3 \pm 1.9$ & $5.6 \pm 2.1$ & $4.2 \pm 2.1$ & $3.1 \pm 2.0$ & $5.4 \pm 2.7$\\
        &Maximum&$10.9 \pm 5.6$&$8.9 \pm 4.5$&$14.8 \pm 4.0$ & $14.1 \pm 4.0$ & $11.9 \pm 3.2$&$10.5 \pm 3.5$&$8.1 \pm 3.6$&$10.2 \pm 3.6$\\
        
        \br
    \end{tabular}\\
    \label{tab:buettner}

\end{table}

    \subsection{3.2 Model Performance for Predicting Parotid Gland Relative Importance}
\begin{figure}[h]
              \centering
              \includegraphics[height=0.6\textheight]{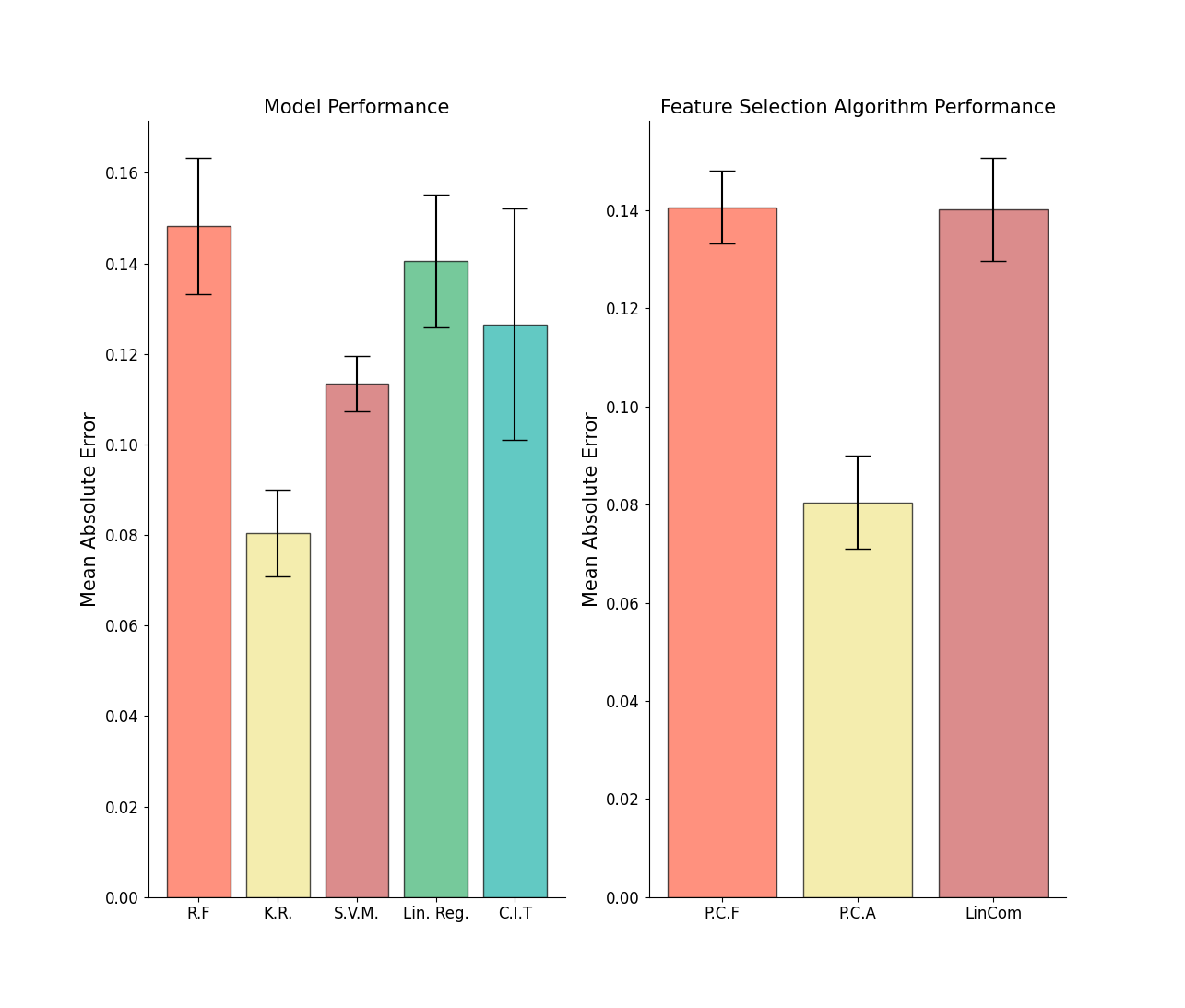}
              \caption{To compare performance across models and feature selection algorithms, the mean absolute error (MAE) of the top-performing model and feature selection algorithm  from each test fold was computed and averaged over all folds. Overall, kernel ridge regression and principal component analysis using 20 features demonstrated the best performance. Error bars correspond to the standard deviation of prediction accuracy over various hyper-parameters.}
              \label{fig:best_models_features}
            \end{figure}

    For each of the nine test-sets of the outer cross validation loop, the MAE, along with best model and feature selection algorithm, as determined via inner cross validation, are shown in Table~\ref{tab:mae_results_all} The average MAE for the radiomics model was 0.08 using enhanced images, and 0.15 with original images. This out-performed the model created using only the mean, median and maximum, which had an MAE of 0.18 and 0.22 for enhanced and original images, respectively. Overall, the best performing model and feature selection algorithm was kernel ridge regression with principal component analysis using 20 features. A performance comparison between all models and feature selection algorithms is shown in Fig~\ref{fig:best_models_features}. The most important features for principal component analysis (determined by projecting principal components scaled by their singular values onto original feature axes), are shown in Table~\ref{tab:top_features}. The overall-best hyper-parameters were found to be a polynomial kernel of degree 2, with $\alpha = 0.1$, and coef0$=1$. $\alpha$ is the regularization strength, and coef0 is the variable-independent coefficient in the kernel function.

        %MAE results Table
        \definecolor{Gray}{rgb}{0.5,0.5,0.5}
        \definecolor{whiter}{rgb}{0.8,0.8,0.8}
        \definecolor{white}{rgb}{1,1,1}
        \newcolumntype{a}{>{\columncolor{Gray}}c}
        \newcolumntype{b}{>{\columncolor{whiter}}c}
        \newcolumntype{d}{>{\columncolor{white}}c}
        \begin{table}[h]
                \centering
                \captionsetup{justification=raggedright}
                \caption{Mean absolute error for each test set of the outer cross validation is shown, along with the best performing model and feature selection algorithm determined during the inner cross validation. Results are shown for the radiomics model and the modle created using only the mean, median, and maximum uptake staitstics. Results are shown for models created with both enhanced and original PSMA PET images.}
                \footnotesize
                \begin{tabular}{@{}dbbbddd}
                \br
                &\multicolumn{6}{l}{Radiomics Model}\\
                \mr
                \rowcolor{white}&\multicolumn{3}{c}{Enhanced Images}&\multicolumn{3}{c}{Original Images}\\
                Fold&M.A.E&Model&F.S Algorithm&M.A.E&Model&F.S Algorithm\\
                1&0.08&K.R&P.C.A&0.31&C.I.T&P.C.A\\
                2&0.10&K.R&P.C.A&0.29&R.F&P.W.C\\
                3&0.06&K.R&P.C.A&0.08&K.R&P.C.A\\
                4&0.04&K.R&P.C.A&0.11&K.R&P.C.A\\
                5&0.06&C.I.T&P.C.A&0.24&R.F&P.C.A\\
                6&0.09&K.R&P.C.A&0.03&C.I.T&P.C.A\\
                7&0.07&K.R&P.C.A&0.02&C.I.T&P.C.A\\
                8&0.13&C.I.T&P.C.A&0.04&K.R&P.C.A\\
                9&0.08&K.R&P.C.A&0.20&C.I.T&P.C.A\\
                \rowcolor{white}Average&0.08&&&0.15\\
                \mr
                &\multicolumn{6}{l}{Mean, Median, Maximum Model}\\
                \mr
                \rowcolor{white}&\multicolumn{3}{c}{Enhanced Images}&\multicolumn{3}{c}{Original Images}\\
                Fold&M.A.E&Model&F.S Algorithm&M.A.E&Model&F.S Algorithm\\
                1&0.08&Lin Reg&N/A&00.27&Lin Reg&N/A\\
                2&0.28&SVM&N/A&0.27&Lin Reg&N/A\\
                3&0.16&SVM&N/A&0.17&Lin Reg&N/A\\
                4&0.03&SVM&N/A&0.09&K.R&N/A\\
                5&0.22&SVM&N/A&0.20&Lin Reg&N/A\\
                6&0.20&SVM&N/A&0.31&K.R&N/A\\
                7&0.12&SVM&N/A&0.16&Lin Reg&N/A\\
                8&0.28&SVM&N/A&0.21&Lin Reg&N/A\\
                9&0.23&SVM&N/A&0.25&Lin Reg&N/A\\
                \rowcolor{white}Average&0.18&&&0.22\\
                \br
            \end{tabular}\\
            \label{tab:mae_results_all}
        
        \end{table}
        
        %IMportant features table
        \begin{table}[h]
            \centering
            \captionsetup{justification=raggedright}
            \caption{Relative importance of radiomic PSMA PET / CT features determined via principal components analysis for modelling of parotid gland subregion importance.}
            \footnotesize
            \begin{tabular}{@{}cccc}
            \br
            &Feature Name&Modality&Relative Importance\\
            \mr
            1.&Original GLCM - Inverse Difference&PSMA PET&1.0\\
            2.&Square Root GLCM - Run Variance&PSMA PET&0.98\\
            3.&Original GLCM - Inverse Difference Moment&PSMA PET&0.97\\
            4.&Square Root GLCM Inverse Difference&PSMA PET&0.97\\
            5.&Original GLCM Inverse Variance&PSMA PET&0.96\\
            6.&Square Root GLCM Inverse Difference Moment&PSMA PET&0.95\\
            7.&Square Root First Order RMS&CT&0.95\\
            8.&Square GLSZM GLNUN&PSMA PET&0.94\\
            9.&Original GLRLM Long Run Emphasis&PSMA PET&0.94\\
            10.&Original GLDM Large Dependence&PSMA PET&0.93\\
            11.&Original GLRLM Run Variance&PSMA PET&0.91\\
            12.&Wavelet HHH GLCM Joint Average&CT&0.91\\
            13.&Wavelet GLCM Sum Average&CT&0.91\\
            14.&Wavelet HHH GLSZM High Gray Level Zone Emphasis&CT&0.91\\
            15.&Square Root GLCM Inverse Variance&CT&0.90\\
            16.&Wavelet HHH GLDM High Gray Level Emphasis&CT&0.90\\
            17.&Wavelet HHH GLRLM High Gray Level Run Emphasis&CT&0.90\\
            18.&Wavelet LLL GLCM Dependence Entropy&CT&0.90\\
            19.&Wavelet HHH GLCM Autocorrelation&CT&0.89\\
            20.&Square Root First Order Entropy&CT&0.88\\

            \br
        \end{tabular}\\
        \label{tab:top_features}

        \end{table}
        %model / fs performance fig

\begin{figure}[h]
              \centering
              \includegraphics[width=0.9\textwidth]{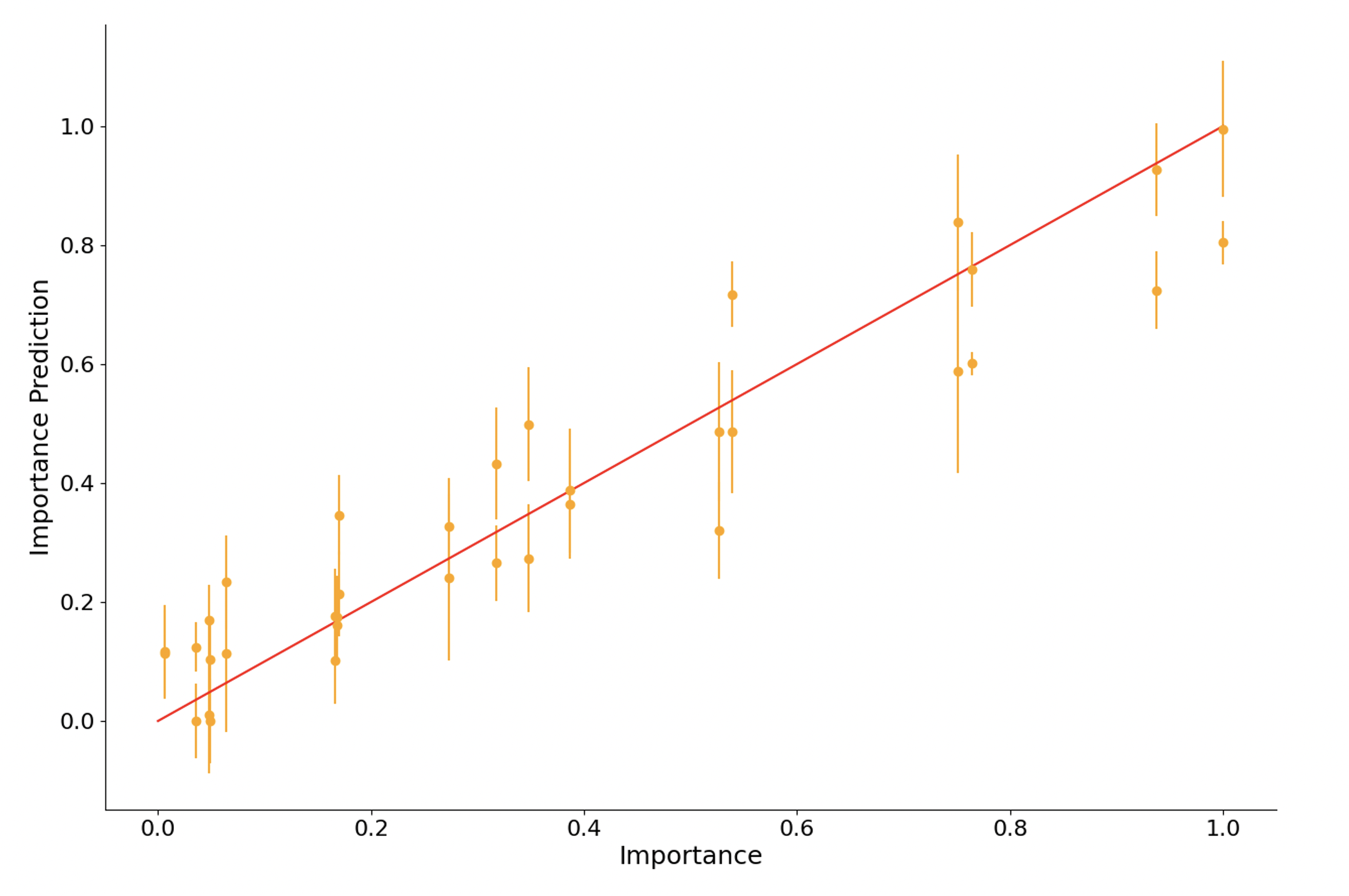}
              \caption{Model-predicted relative importance estimates for all subregions are plotted along with Clark et al.'s (in this case, ground truth) importance estimates. Predictions are shown with their associated model predicted errors. Predictions were made using the best performing model and feature selection algorithm found with nested cross validation - a kernel ridge regressor model and principal component analysis feature selection using 20 features.}
              \label{fig:test_preds}
        \end{figure}

        All model predictions for importance in the 9 test sets are collected and plotted against Clark's estimates in Fig~\ref{fig:test_preds}. Prediction error was estimated by a separate kernel ridge model which was trained to predict model estimation error, as previously described.
%model / fs performance fig

    \subsection{3.3 Estimating Deviations of Patient-Specific Importance from population Estimates}
\begin{figure}[h]
              \centering
              \includegraphics[width=1\textwidth]{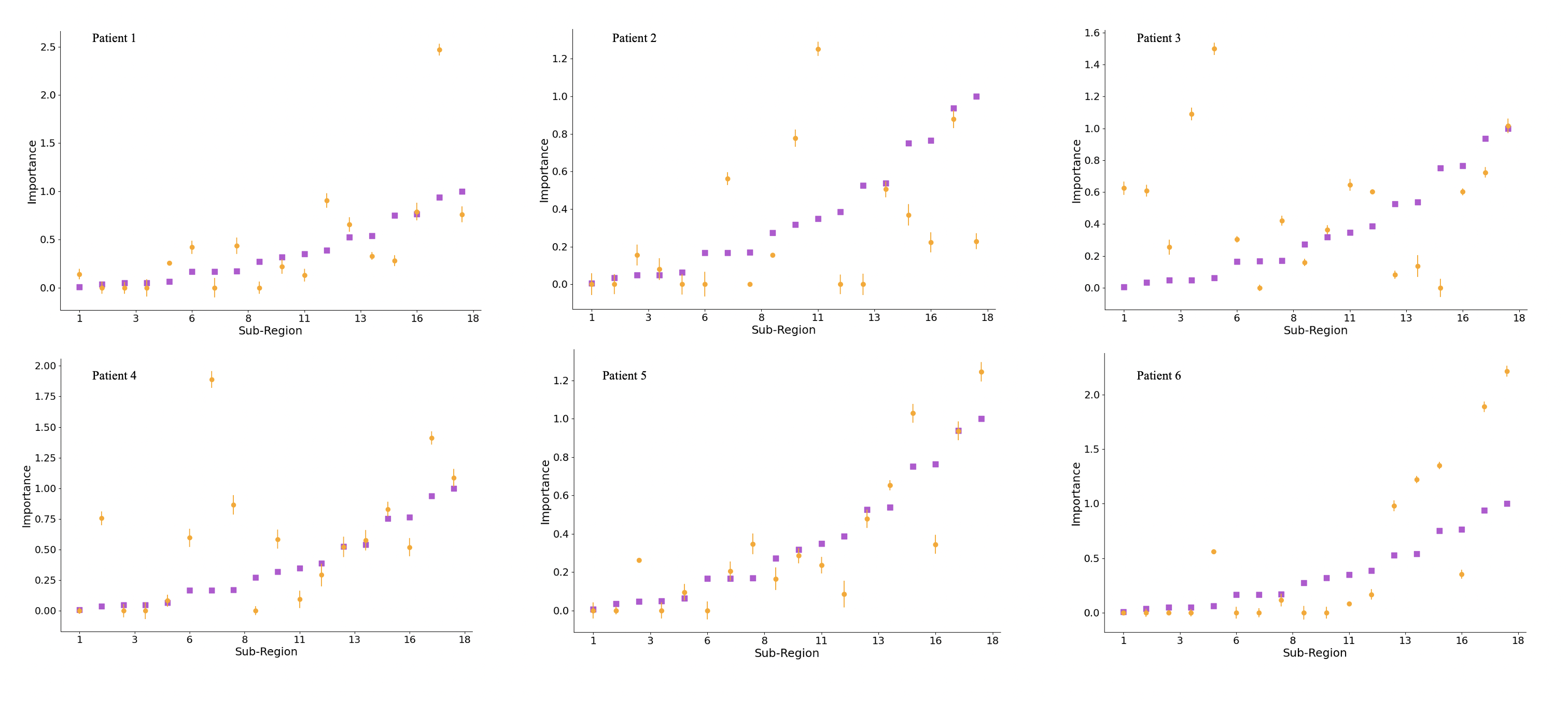}
              \caption{Importance estimates obtained for six individual patients using the population-level model, are shown. Population level estimates for the 18 subregions are shown as purple squares, with the patient specific estimates as gold circles. The population-level model captures the relationship between important radiomic features and importance estimates, and can be used for estimating approximate shifts in importance estimates for an individual patient's parotid gland subregions. Error estimates were obtained with the kernel ridge error model.}
              \label{fig:patient_preds}
        \end{figure}

        The best performing model, feature selection algorithm, and hyper-parameters as determined via cross validation, were then used to train a final population-level model using the entire population-level data set. Importance estimates for parotid gland subregions of individual patients could then be obtained by inputting a given patient's radiomic features into the model. Examples of how individual predictions deviate from population-level estimates, along with prediction errors estimated with the kernel ridge error model, are demonstrated for six different patients in Fig~\ref{fig:patient_preds}. 
    
\begin{figure}
              \centering
              \includegraphics[width=0.8\textwidth]{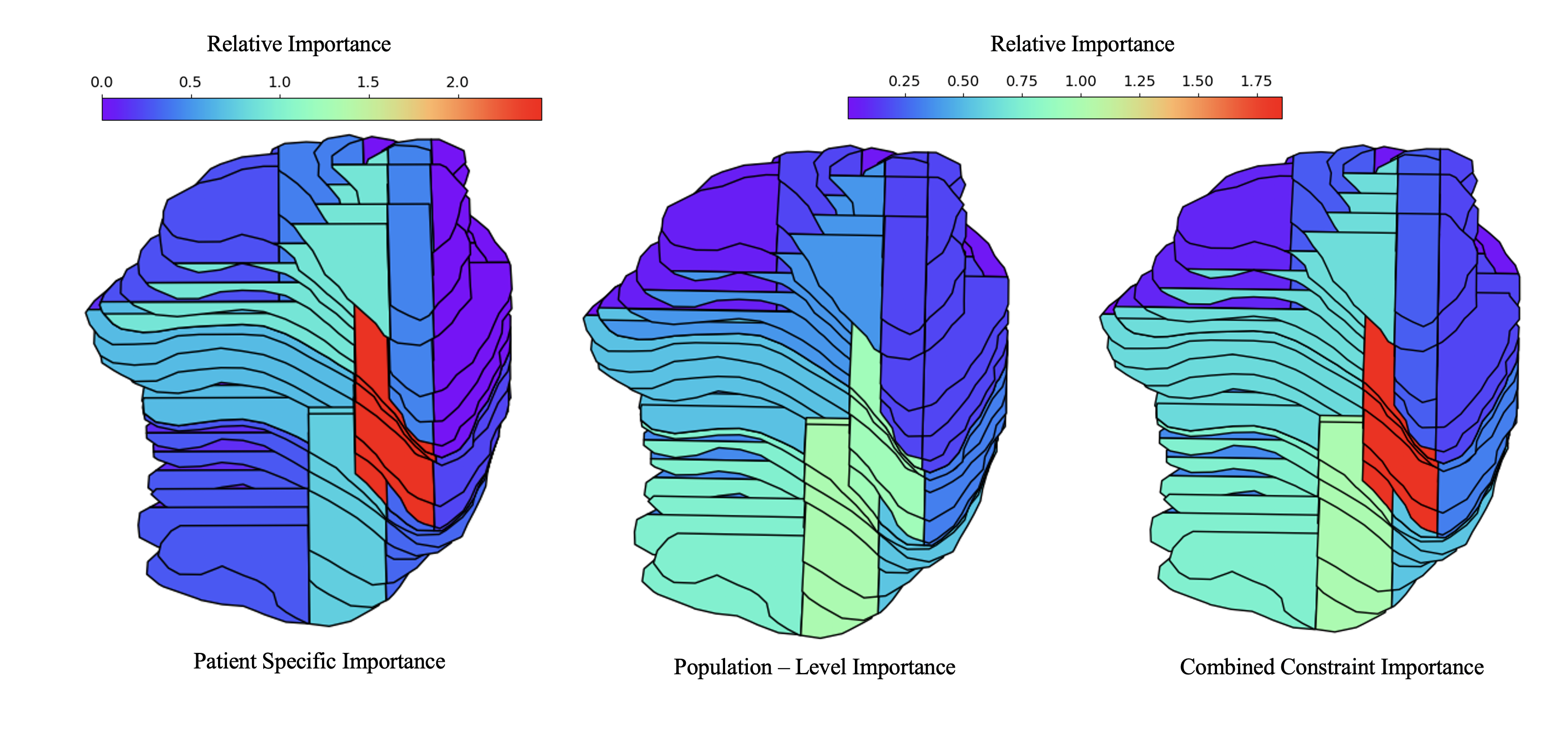}
              \caption{We demonstrate how patient-specific parotid subregional importance estimates (left), can be used to supplement population-level importance estimates (middle) using the formula described in the methods, such that final estimates are never lower than population-level estimates, but further increased in regions where patient-specific estimates are high. Only positive perturbations to regional importance were made, to avoid negatively impacting patients in the case where importance estimates are to be used for designing dose constraints. }
              \label{fig:perturbations}
        \end{figure}

        Fig~\ref{fig:perturbations} illustrates an example of how an individual patient's parotid gland radiomic features can be used to supplement the population importance estimate, using the equation described in the methods.

\afterpage{\clearpage}
\newpage
\section{Discussion:}
    The results of this work indicate that intra-parotid gland PSMA PET uptake may be inversely related to subregional relative functional importance. Anticorrelative relationships were found to exist between PSMA PET uptake and four independent models of subregional parotid gland importance from the literature \cite{clark_2018, han, vanluijk_2015, buettner_2012}. These findings are interesting in relation to recent findings that suggest a direct relationship between inter-parotid gland PSMA PET uptake and whole-gland functionality \cite{zhao, mohan, hotta, Mohan2022}. While it appears from previous studies that whole-parotid gland functionality is directly proportional to the magnitude of PSMA PET uptake, these results suggest that subregions of high relative functional importance within glands tend to have lower uptake. It is unclear which physiological mechanisms these findings correspond to.
    
    Based on the present findings, PSMA PET may be a useful quantitative imaging modality for assessing intra-parotid gland functionality. As xerostomia remains a common burden for head and neck radiotherapy patients \cite{ma}, deeper understanding of intra-parotid gland functionality is important for enabling modernized parotid gland dose constraints to be designed. 
    
    Clark et al.'s \cite{clark_2018} and Han et al.'s \cite{han} relative importance estimates were practically advantageous in that they define numerous, well-defined, and non-overlapping subregions of parotid glands where correlations between relative importance and PSMA PET uptake could be assessed. Both models predict higher importance towards medial and caudal-middle regions of the gland, while Clark et al.'s predicts much higher importance in the anterior half of the gland. Clark et al.'s model was significantly ($p < 0.02$) anti-correlated with PSMA PET uptake, while Han et al.'s model was also anti-correlated, but not significantly. It should be noted, however, that regions of high importance predicted by both models correspond to regions of lower than average uptake \cite{sample_2023_hetero}. Han et al.'s sub-segmentation method yields subregions of unequal volume, which also creates problems when comparing uptake statistics. Due to the shape of the parotid glands, subregions within central regions will be larger than those at the superior and especially the inferior portion when segmention yields subregions of equal superior-inferior length. Both Clark et al.'s and Han et al.'s models were developed using stimulated salivary measurements. 

    We were able to approximate the location of Van Luijk et al.'s \cite{vanluijk_2015} critical regions and compute PSMA PET uptake statistics within and outside said regions. Uptake statistics within critical regions were significantly lower than non-critical regions, corroborating the anticorrelative trend between importance and PSMA PET uptake observed with Clark et al. and Han et al.'s subregions. Sparing dose in Van Luijk' et al.'s \cite{vanluijk_2015} critical regions of parotid glands for radiotherapy patients was recently shown to insignificantly impact patient outcomes \cite{steenbakkers}. However, dose to critical regions was more predictive of salivary dysfunction than whole-gland dose. Comparing uptake with importance in regions defined by Buettner et al.'s \cite{buettner_2012} analysis also pointed towards an anticorrelative relation of importance with PSMA PET uptake.
    
     Simultaneous deblurring and supersampling of PSMA PET images \cite{sample_2023_blind_deconv} prior to uptake calculations led to stronger correlations between uptake and importance estimates, and better model performance for predicting regional importance. Better performance using enhanced images was expected, as partial volume effects cause fine detail wash-out in small regions of PET images.

    The relationship between PSMA PET and relative importance appears non-linear (Fig~\ref{fig:Clark_Scatter}) and proved to be better predicted using radiomic features and non-linear modelling (Fig~\ref{fig:test_preds}. Model development for predicting relative importance with PSMA PET and CT radiomic features was successful, yielding a relatively low MAE (0.08) for test predictions, considering the small size of the data set. Kernel ridge regression with a quadratic kernel out-performed all other models tested. This suggests that the relationship between PSMA PET uptake and regional importance is not simply linear. This is further supported by the most important features determined by principal component analysis (Table~\ref{tab:top_features}). A relatively high number of features (20) was found to perform best on tests sets of cross validation. This number was a hyperparameter of training, and many smaller values were compared (as well as one larger value) as indicated in Table~\ref{tab:models_and_fs_tried}. 

    Using radiomic features of PSMA / PET and CT images was advantageous over using only the mean, median, and maximum uptake statistics for model-building to predict regional importance. Radiomic features of squared PSMA PET uptake were found to be particularly predictive of subregional parotid gland importance. The top six most important features were all forms of the GLCM of PSMA PET images. Based on this finding, we recommend including radiomics of squared uptake in future predictive models of functional importance for salivary glands. 
    
    A method of using PSMA PET and CT radiomic features to predict patient deviations from population-level estimates of parotid gland regional importance was demonstrated. The purpose was to present a hypothetical method of extracting patient-specific parotid gland importance estimates for tailoring patient dose constraints for radiotherapy \cite{sample_2021}. PSMA PET is not acquired as a standard-of-care image for head-and-neck cancer patients, and it is likely cost-prohibitive to add PSMA PET to the standard-of-care. Therefore PSMA PET images would not be practically available in most clinical situations. However, we believe further PSMA PET studies could be critical in shedding light on importance trends and variability within the parotid glands.

    The incidence of head and neck cancers in patients under the age of 45 has increased sharply in the last two decades due to a rise in oropharyngeal cancers caused by the human papillomavirus (HPV) \cite{Sturgis2007, Mehanna2012}, and xerostomia remains a common side effect for head-and-neck cancer patients \cite{ma}. Xerostomia leads to significant reductions in self-assessed quality-of-life scores \cite{cassolato2003xerostomia, pedersen2002saliva, ekstrom2019saliva, villa2011dental}, and is largely a result of radiation received in salivary glands during radiotherapy. Dose to the the parotid glands is the greatest risk factor for post-treatment xerostomia \cite{chambers}. The current standard of care is to minimize the whole-mean dose to parotid glands \cite{deasy}, despite evidence of intra-gland functional variability \cite{clark_2018, han, buettner_2012, vanluijk_2015}. Advancements in the understanding of intra-parotid gland functional heterogeneity and the ability to predict relative subregional importance using PSMA PET has the potential application of creating and testing new salivary gland dose constraints for radiotherapy treatment planning.

    The results of this study support the potential utility of using PSMA PET/CT for designing relative-importance based dose constraints that can be validated in clinical studies. Radiomic features appear to be particularly useful for predicting importance, and will likely play a central role in future studies which aim to develop population-level or patient-specific dose constraints whose efficacy can be validated in a clinical study.

    Our data set was small, comprised of only 60 parotid glands from 30 patients. This necessitated the double cross validation methodology used for model development, where the test set was rotated through, along with inner validation sets for each, to determine model parameters. Outer test sets had no influence on model development and were used to independently test the models predictive accuracy of all subregions. The majority of Clark et al.'s \cite{clark_2018} subregions have low relative importance, with only a few regions having high relative importance. As a result of this behaviour, double cross-validation was warranted to help mitigate biased error estimates when validating with high importance values that were not adequately represented in the training set. As all PSMA PET/CT images in this dataset were acquired on a single scanner, top performing parameters found in this study should not necessarily be expected to match those found in future model-building studies using external datasets.

    \section{Conclusion}
    The results of this work, which compared four models of parotid gland subregional importance from the literature with regional PSMA PET uptake, demonstrate an inverse proportionality between relative importance and subregional PSMA PET uptake. We demonstrated the utility of PSMA PET radiomic features for predicting regional importance by building a predictive model for Clark et al.'s regional importance estimates (MAE = 0.08). Lastly, we demonstrated a methodology for supplementing population-level importance estimates using patient-specific radiomic features. PSMA PET appears to be a promising quantitative imaging modality for analyzing salivary gland functionality.

\section*{Acknowledgments}
This work was supported by the Canadian Institutes of Health Research (CIHR) Project Grant PJT-162216. 

\section*{Ethical Statement}
This retrospective study was approved by the BC Cancer Agency Research Ethics Board (H21-00518-A001). Participants had provided written consent in prior studies for their data to be further analyzed and for subsequent results to be published.

\section*{References}

\bibliography{bib} 
\bibliographystyle{ieeetr}

\end{document}